\def\erg{{\rm\thinspace erg}}

\def\K{{\rm\thinspace K}}

\def\km{{\rm\thinspace km}}
\def\kpc{{\rm\thinspace kpc}}

\def\Mpc{{\rm\thinspace Mpc}}
\def\Msun{\hbox{$\rm\thinspace M_{\odot}$}}

\def\s{{\rm\thinspace s}}
\def\yr{{\rm\thinspace yr}}

\def\ergpMpc{\hbox{$\erg\Mpc^{-3}\,$}}
\def\ergps{\hbox{$\erg\s^{-1}\,$}}

\def\kmps{\hbox{$\km\s^{-1}\,$}}

\def\Msunpyr{\hbox{$\Msun\yr^{-1}\,$}}

\def\etal{\hbox{et al.}}

\documentclass{rspublic}
\usepackage{psfig}
\usepackage{graphicx}
\usepackage[authoryear]{natbib}
\begin{document}

\title[AGN and the ICM]{Active Galaxies and Cluster Gas}

\author{A.C Fabian}
\affiliation{Institute of Astronomy, Madingley Road, Cambridge CB3 0HA, UK}
\label{firstpage}
\maketitle

\begin{abstract}{X-rays; active galaxies; clusters; cooling flows}

Two lines of evidence indicate that active galaxies, principally radio
galaxies, have heated the diffuse hot gas in clusters. The first is
the general need for additional heating to explain the steepness of
the X-ray luminosity--temperature relation in clusters, the second is
to solve the cooling flow problem in cluster cores. The inner core of
many clusters is radiating energy as X-rays on a timescale much
shorter than its likely age. Although the temperature in this region
drops by a factor of about 3 from that of the surrounding gas, little
evidence is found for gas much cooler than that. Some form of heating
appears to be taking place, probably by energy transported outward
from the central accreting black hole or radio source. How that energy
heats the gas depends on poorly understood transport properties
(conductivity and viscosity) of the intracluster medium.
Viscous heating is discussed as a possibility. Such heating processes have
consequences for the truncation of the luminosity function of massive
galaxies.

\end{abstract}

\section{Introduction}

Clusters of galaxies are luminous X-ray sources, with X-ray
luminosities $L_{\rm X}\sim 10^{43}- 10^{46}\ergps$. The emission is
predominantly thermal bremsstrahlung from highly ionized hydrogen and
helium in the intracluster medium (ICM) at temperatures $T\sim
10^7-1.5\times 10^8\K$. Line emission, particularly from iron, is also
present showing that most of the gas has a mean metallicity of about
0.3 Solar. The total mass of the intracluster medium is about one
sixth of the total cluster mass, and the stars in all the member
galaxies have about one sixth of the mass of the hot gas. Most of the
total mass of a cluster is due to dark matter.

Simple gravitational collapse models for clusters scale such that the
$L_{\rm X}\propto T^2$ whereas the data are better fit by $L_{\rm
X}\propto T^3$ or even steeper at low ICM temperatures.  This can be
explained by additional heating of the ICM to a level of 1--3~keV per
particle. Supernovae are unable to supply this much heat and active
galactic nuclei (AGN) within a cluster, or its consitutent subclusters
at an earlier epoch, seem the most likely heat source on energetic
grounds (Valageas \& Silk 1999; Wu, Fabian \& Nulsen 2000).

The X-ray emission in the cores of many clusters is sharply peaked.
The radiative cooling time of the gas within 50~kpc of the centre is
shorter than the likely age of the cluster. The temperature drops
smoothly there by a factor of two to three from that of the outer gas.
Although it might seem that a cooling flow should be operating there
with gas cooling out of the intracluster medium, spectra from
XMM-Newton and Chandra show that radiative cooling is much reduced and
some form of distributed heating is taking place.

This last point is of considerable importance for understanding the
gaseous part of galaxy formation, much of which proceeds by radiative
cooling of hot gas in dark matter potential wells. The cooling
in galaxies predominately occurs in the extreme and far UV and so is
not readily observable, but is directly observable in clusters.
Whatever is stemming the cooling in clusters may be determining the
upper mass cutoff for galaxies.

The AGN in the central galaxy is a viable culprit for heating
the core region of clusters. AGN may therefore play a significant
role in determining the properties of the ICM, both generally for the
whole cluster and specifically for the inner region. 

\section{Evidence for AGN heating of clusters at moderate to high redshifts}

Massive galaxies host massive black holes built by accretion over the
redshift range 2--4. Many of them plausibly were powerful radio
galaxies at some stage, injecting vast amounts of energy into
surrounding gas. Integrating the radio galaxy luminosity function over
time leads to a possible comoving energy input of
$\sim10^{57}\ergpMpc$ (Inoue \& Sasaki 2001). This can preheat the ICM
to 1--2~keV per particle, as required to account for the observed
$L_{\rm X}\propto T^3$ rather than $T^2$ as expected from gravity
alone.

Studying this AGN--cluster connection further is not easy. We need to
know the kinetic power of jets and their lifetime. Radio observations
of course show us that such sources exist but synchrotron and Compton
losses on the electrons mean that the observable lifetime is short. One
recent development which should help is the detection with Chandra of
extended X-ray emission from high redshift radio galaxies. The
emission is likely inverse Compton scattering of Cosmic Microwave
Background (CMB) photons by electrons with Lorentz factor $\gamma\sim
1000$. The steep increase in the energy density of the CMB with
redshift, as $(1+z)^4$, partially compensates for the large distance
to the sources, so making them observable. By $z\sim2,$ $L_{\rm X}$ for
a given electron population is boosted by a factor of nearly 100
compared with the present epoch.

\begin{figure}
\begin{center}
\includegraphics[width=0.5\textwidth,angle=0]{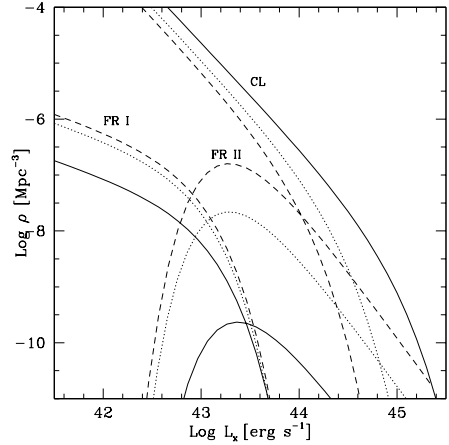}
\caption{X-ray luminosity function of clusters (CL) and radio galaxies
adapted from the 151~MHz luminosity function (Willott et al 2001)
assuming $\nu_{\rm R}L_{\rm R}=\nu_{\rm X}L_{\rm X}$ (Celotti \&
Fabian 2004). Solid, dotted and dashed lines refer to $z=0$, 1 and 2.}
\end{center}
\end{figure}

Examples of such sources are 3C294 ($z=1.786$, Fabian et al 2003c), 3C9
($z=2.01$, Fabian, Celotti \& Johnstone 2003), PKS\,1138-262 (Carilli
et al 2002), 4C41.17 ($z=3.8$, Scharf et al 2003) and GB1508+5714
($z=4.3$, Yuan et al 2003; Siemiginowska et al 2003). As discussed by
Celotti \& Fabian (2004), similar sources should emerge from further
Chandra observations of distant radio galaxies. They should dominate
the extended sources in the X-ray Sky at $z>2$ and $L_{\rm
X}>10^{44}\ergps$ (Fig.~1). Since the X-ray emission involves lower energy
electrons than typically involved in the observed radio emission, a
source should be detectable longer after an outburst in the X-ray band
than in the radio. (Note that the X-ray extent of 3C294 exceeds the
radio emission by $\sim50\kpc$ or more to the NW and SE.)  This means
that some unidentified faint extended X-ray sources (Bauer et al 2002)
could be due to inverse Compton emission from faded radio galaxies.

\section{Cluster cores}

The radiative cooling time within the inner 100~kpc of most cluster
cores is less than $10^{10}\yr$. The gas temperature also drops there
by a factor of about three (Fig.~2). If there is no heating of the gas it
should cool out at a rate given by (see Fabian 1994 for a review)
\begin{equation}
\dot M_{\rm X}={2\over5}{{L\mu m}\over{kT}},
\end {equation}
reduced by a factor of about two by gravitational energy release as
the gas flows. To maintain the pressure required to support the weight
of overlying gas, the cooling gas flows inward, forming a
cooling flow.
 
\begin{figure}[h]
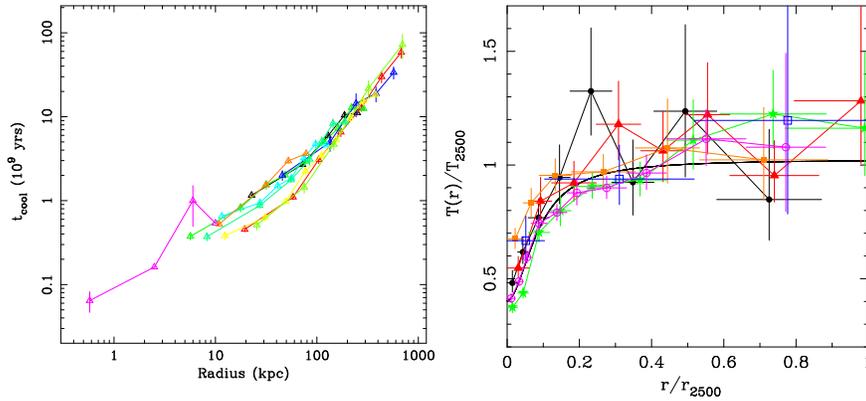

\includegraphics[angle=-90,width=0.45\textwidth]{tcool_all2.ps}
\includegraphics[angle=-90,width=0.45\textwidth]{temp_profile.ps}
\caption{Left: Radiative cooling time versus radius for several cooling flow
clusters (Voigt \& Fabian 2003). Right: Temperature
profile for 6 massive clusters (Allen et al 2001).}
\end{figure}

As the gas cools below 1~keV it emits strong Fe L line emission (e.g.
FeXVII emission at 15 and 17~A). A major result from the Reflection
Grating Spectrometer (RGS) on XMM-Newton was to show that little such
emission is seen (Peterson et al 2001, 2003; Tamura et al 2001; Kaastra
et al 2004; Fig.~3). These studies show that the mass cooling rate
below about one third of the bulk cluster temperature is less than one
fifth to one tenth of that deduced from the above simple
formula. Chandra data give the same result.

\begin{figure}
\begin{center}
\includegraphics[width=0.45\textwidth,angle=-90]{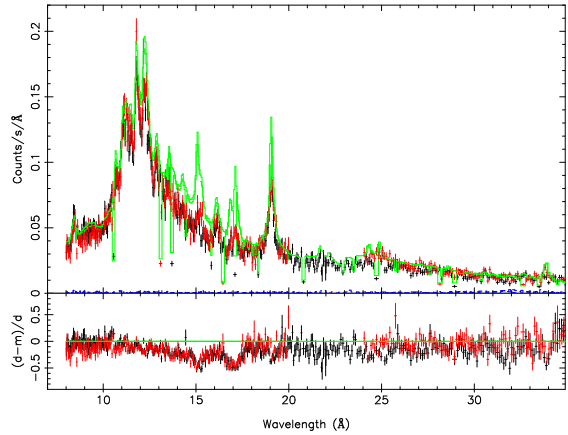}
\caption{RGS X-ray spectrum of the inner parts of the Virgo cluster
around M87 (Sakelliou et al 2002). The faint line shows the predicted
emission if gas is cooling to below $10^6\K$. The data clearly shows
much less emission than predicted by this model in the 13--18~A region. }
\end{center}
\end{figure}
\begin{figure}
\includegraphics[width=0.4\textwidth,angle=-90]{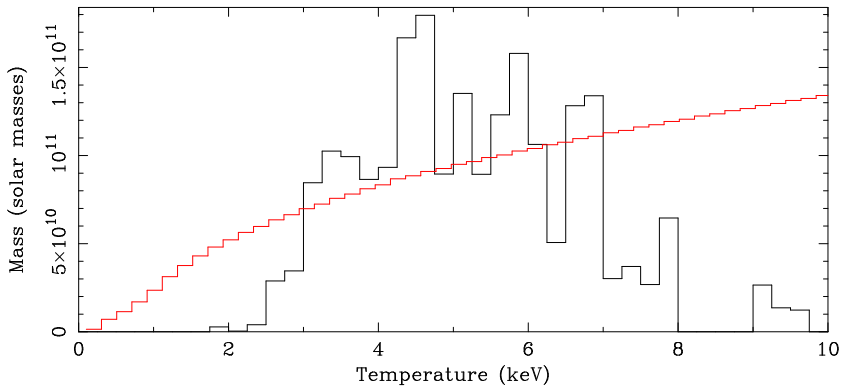}
\caption{Mass -- temperature distribution in the core of the Perseus cluster
(Sanders et al 2004). Gas is seen down to about 2.5~keV. The curved
line shows the expected mass distribution from gas cooling at constant
pressure from the (outer) virial temperature (7~keV for Perseus) to
below 0.1~keV. Gravitational work done as the gas flows, and isochoric
cooling if magnetic pressure dominates, would reduce the low
temperature prediction by a factor of up to 3. Nevertheless no gas is
seen below 1.5~keV. }
\end{figure}

The effect was present in previous ASCA and ROSAT studies (e.g., Allen
et al 2001) where the lack of emission had been attributed to
intrinsic absorption. Although some absorption has not been completely
ruled out, the improved new spectra show that it is either not
dominant or intrinsic to the coolest gas.

The gas in a cluster core appears to cool by about a factor of three
from the cluster virial temperature but no further. Since $t_{\rm
cool}\propto T^2$ at constant pressure around 1~keV, the cooler
the gas the faster it cools. Why it does not cool further is
puzzling and constitutes the 'cooling flow problem'. Just rearranging
the cooling gas or only heating the coolest gas does not solve this
problem. The data (Fig.~4) show no accumulation of gas at any
particular temperature, just a lack of gas below about $1/3$ of the
virial temperature.

It is likely that some heating is taking place, for which there are
two plausible, long-known, candidates. They are heating by conduction
from the hot outer gas and heating by a central active nucleus (Tucker
\& Rosner 1983; Rosner \& Tucker 1989; Tabor \& Binney 1993; Binney \&
Tabor 1995). However, a range of one-dimensional heating models
investigated by Brighenti \& Mathews (2003) and by Ruszkowski \&
Begelman (2002) shows that the solution of the cooling flow problem is
not trivial.

\subsection{Thermal conduction}

Thermal conduction has long been considered to be suppressed in
clusters because of the observed central temperature drop. Conductive
energy flow increases strongly with temperature, unlike radiative
cooling which decreases (at constant pressure), so one might assume
that it either operates, so making the core isothermal, or is
suppressed and radiative cooling dominates. Narayan \& Medvedev (2001)
have however revived the concept and noted that conduction may account
for the observed temperature gradients. This has been explored in more
detail by Fabian et al (2002b), Voigt et al (2002; Fig.~5) and Voigt \&
Fabian (2004). These last authors found some clusters where conduction
appears to be insufficient, therefore ruling out simple electron
conduction models. A further issue is whether the effective
conductivity can even be as high as the Spitzer value, or whether
magnetic fields inferred from Faraday rotation (Fig.~6) suppress it
heavily. (Note also that conduction models do not explain where the large
observed temperature drop comes from in the first place.)

\begin{figure}[ht]
\begin{center}
\includegraphics[width=0.45\textwidth,angle=-90]{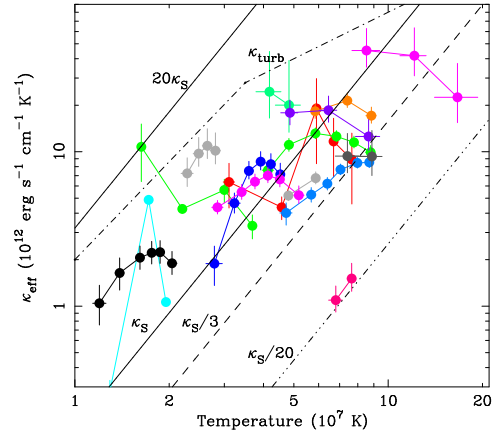}
\caption{Effective conductivity required to balance radiative
cooling in a sample of the brightest clusters (Voigt \& Fabian
2004). The Spitzer rate is labelled $\kappa_{\rm S}$.  }
\end{center}
\end{figure}

Cho et al (2003) have pointed out that turbulent heat diffusion may be
important in cluster cores. Lisa Voigt and I (2004) have found that it
is of the right form and magnitude ($\kappa_{\rm turb}$ in Fig.~5) to
explain the observations if the diffusion coefficient is about
$0.2-0.5$ times the sounds speed times the radius (see also Kim \&
Narayan 2003). Such a model assumes that large scale turbulent motions
of gas enable efficient heat exchange within the core. Note that
radial flows require energy, given the outward increasing entropy
gradient, and moreover that abundance gradients (Fig.~6) must be maintained.

\begin{figure}
\includegraphics[width=0.5\textwidth,angle=0]{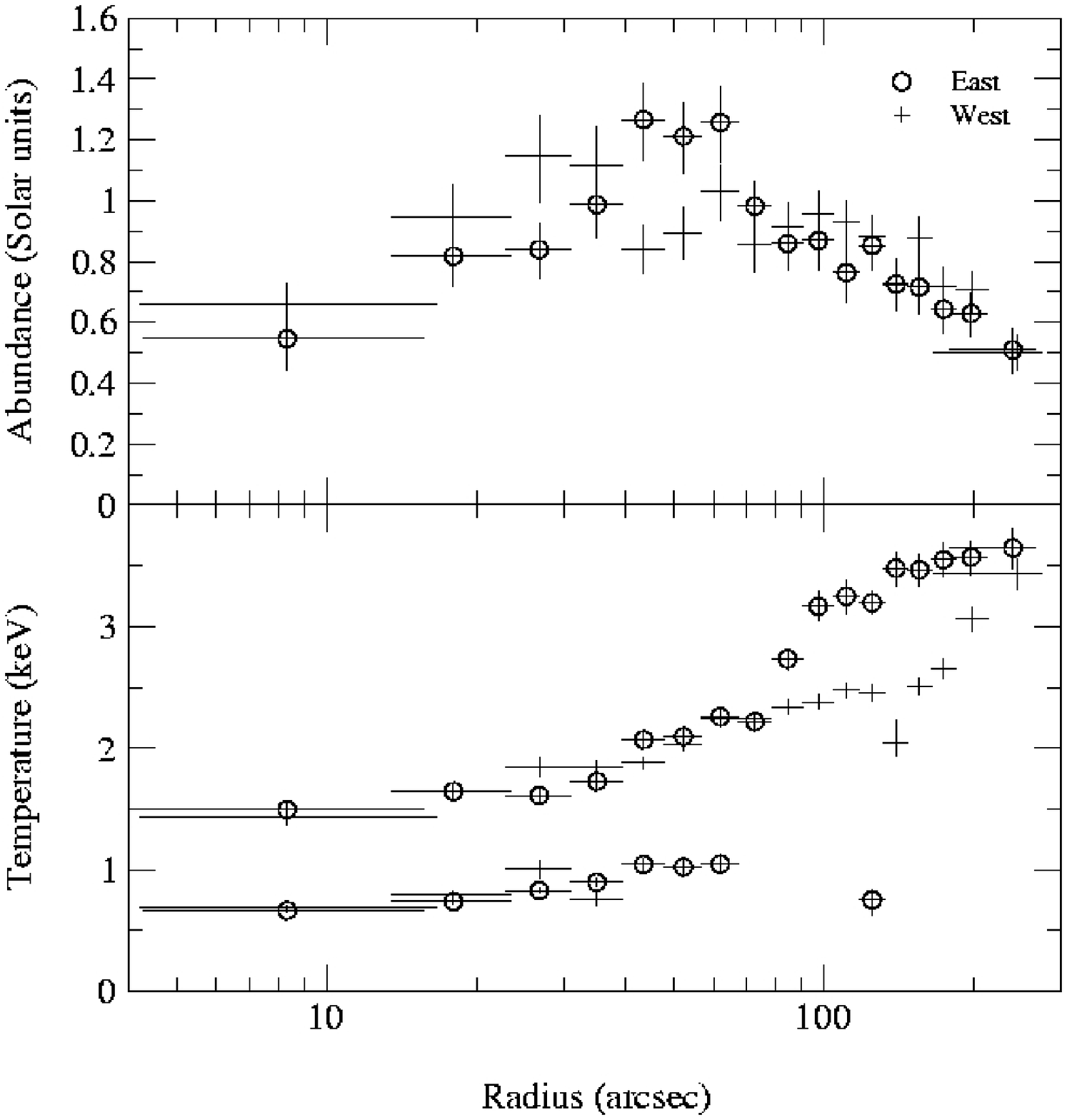}
\includegraphics[width=0.5\textwidth,angle=0]{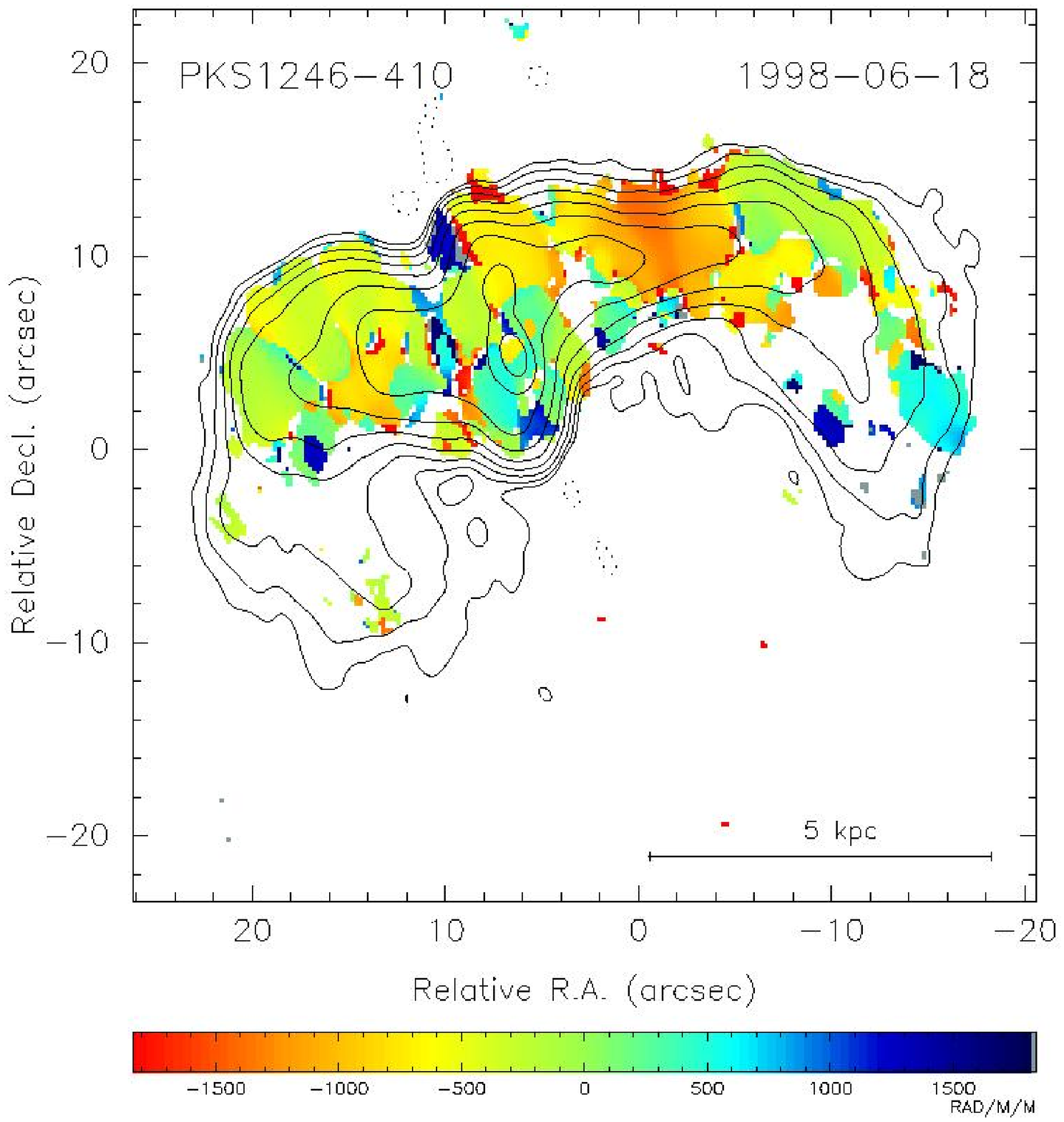}
\caption{Left: Abundance (top) and temperature (bottom) profiles for
the Centaurus cluster (Sanders \& Fabian 2002). Right: Faraday
rotation measure map for the Centaurus cluster (Taylor et al 2002).}
\end{figure}

\begin{figure}[h]
\includegraphics[width=0.5\textwidth]{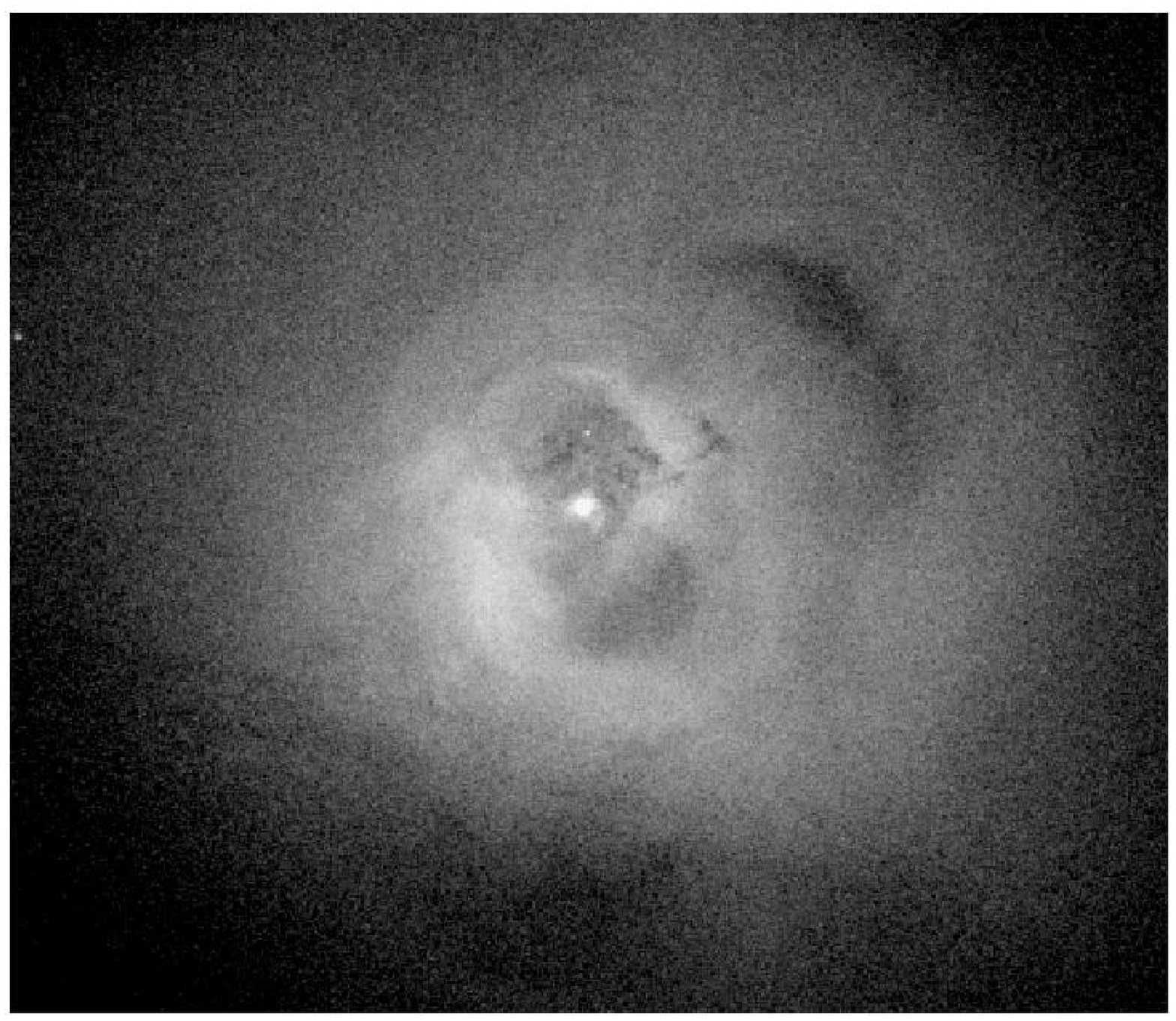}
\includegraphics[width=0.5\textwidth]{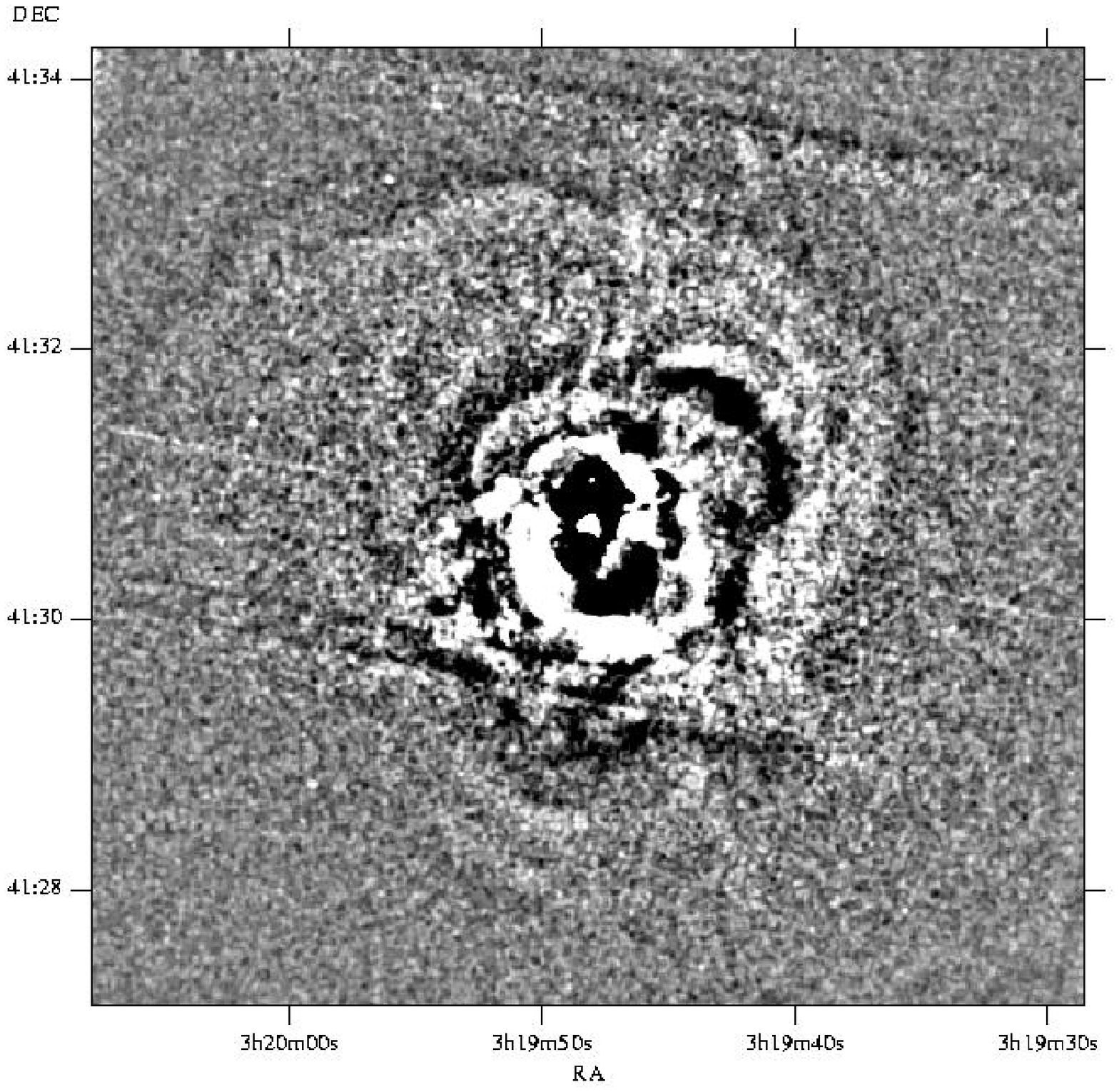}
\caption{Left: X-ray image of the core of the
Perseus cluster. Note the holes, which coincide with the radio lobes,
above and below the nucleus. A buoyant outer bubble is seen to the
right (Fabian et al 2003a). Right: Unsharp-masked X-ray image showing 
the ripples (Fabian et al 2003a).}
\end{figure}

\begin{figure}
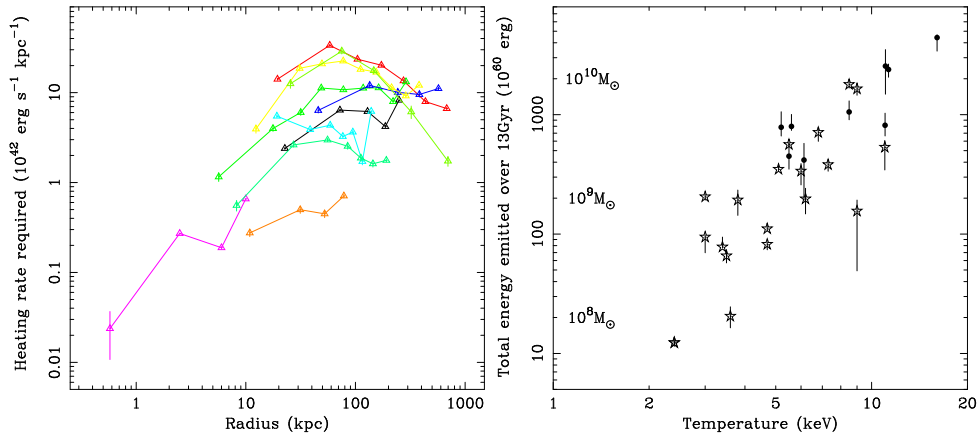

\includegraphics[width=0.45\textwidth,angle=-90]{rate_all.ps}
\includegraphics[width=0.45\textwidth,angle=-90]{cluster_heat.ps}
\caption{Left: The heating rate per unit radius required to stop
radiative cooling in several cooling flow clusters (Voigt \& Fabian
2003). The heating needs to be distributed. Right: The total energy
radiated over a Hubble time from within the cooling region for a
selection of cooling flow clusters (Fabian et al 2002b). The masses
indicate the total mass which must accrete to produce this energy if
the efficiency of energy release is 0.1; the central black hole must
therefore be more massive than this. If the central radio source stems
cooling in the hotter clusters then most of the power released must be
channeled into heating the intracluster medium. }
\end{figure}

\subsection{The central AGN}

All of the relevant {\it cooling flow} clusters peak on a central
galaxy which is expected to host a massive black hole. Many of these
galaxies have radio sources, some of which are obviously blowing
bubbles of relativistic plasma in the central regions (e.g., Perseus, Fabian
et al 2000, 2003, Fig.~7; A2052, Blanton et al 2001; A2597, McNamara
et al 2001; A4059, Heinz et al 2002). The energy flux from the radio
source can be high ($10^{43-45}\ergps$); the jet power can be
determined from the age and size of the bubbles (Fabian et al 2002).

The heat must be distributed (Johnstone et al 2002, Voigt \& Fabian
2003; Fig.~8) and cannot just heat the innermost, coolest gas (Voigt
\& Fabian 2003). Some (Binney \& Tabor 1995; Kaiser \& Binney 2003)
have argued that radio source activity may be sporadic, which explains
why there is little correlation between the present radio source
activity and the heating requirement. For a large duty cycle this
makes very strong demands on the power of the source when it is
switched on, particularly in the high luminosity clusters (Fig.~8),
where it must exceed $10^{46}\ergps.$ This will hardly be contained
in simple bubbles, but will expand and dump its energy into the outer
cluster.

Strong abundance gradients are found in many clusters (Fukazawa et al
2004), often peaking at radii $\sim 30\kpc$ (Johnstone et al 2002;
Fig.~6, Sanders \& Fabian 2002). This limits the degree of large scale
disturbance taking place in the central regions. The apparent central
drop in abundance is puzzling (if not due to resonance scattering).

Further difficulties are that the X-ray coldest observed gas lies
around the bubbles, not all clusters host powerful enough radio
sources. The mechanism for heat transfer from bubbles to the
surrounding gas has been unclear, despite several impressive
computational studies (e.g. Br\"uggen \& Kaiser 2002, Quilis et al
2002; Br\"uggen 2003; Basson \& Alexander 2003; Robinson et al 2003;
Omma et al 2004) and analytical work (Churazov et al 2001; Soker et al
2003). None of the early simulations makes bubbles
resembling those observed.

The creation of the bubbles or cavities means that $PdV$ work is done
on the surrounding gas. This will create sound waves which can
transport that energy away. The sound waves are indeed observed in our
deep image of the Perseus cluster (Fig.~7; Fabian et al 2003a) as ripples
in the X-ray surface brightness. The wavelength is about 10~kpc which,
given the sound speed of the gas, means a period of about $10^7\yr$.
This agrees with the timescale on which bubbles would be formed and
then buoyantly detach from a steady jet at the centre (Churazov et al
2001). Of
course it does not need to be exactly steady but it could be so.

Interestingly we have found that the sound waves will dissipate their
energy over the cooling region if the (ion) viscosity of the gas is
close to the Spitzer-Braginskii value for an unmagnetized plasma
(Spitzer 1962). This would provide a splendid way for the energy
produced by accretion onto the central black hole to be dissipated in
a general, isostropic manner in the surrounding hot gas. Normally,
energy from accretion does not couple well with hot gas since if
electromagnetic the gas is transparent to the radiation and if kinetic
the power is generally highly collimated.

\begin{figure}[h]
\begin{center}
\includegraphics[width=0.6\textwidth]{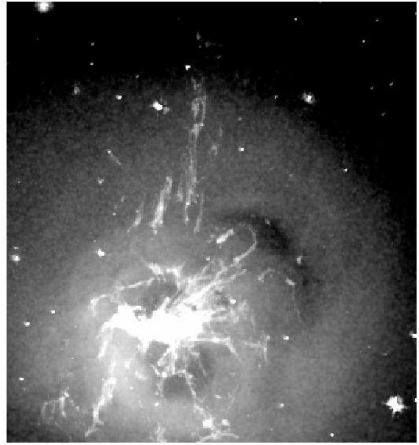}
\caption{The H$\alpha$ image (Conselice et al 2000) superimposed on the X-ray image 
of the centre of the Perseus cluster (Fabian et al 2003b).  Note the
many long straight outer filaments and the horseshoe ones inward of
the outer bubble. }
\end{center}
\end{figure}

\subsection{The viscosity of the ICM}

We have some independent evidence that the viscosity is high from the
H$\alpha$-emitting optical filaments seen (e.g. Conselice et al 2000)
in the Perseus core around the central galaxy NGC\,1275. If the
viscosity is low then the hot gas should be turbulent, yet the
filaments, which are less than a parcsec thick as estimated from their
brightness, are often very straight for tens of kpc (Fig.~9).
Moreover, just within the outer bubble there are two
'horseshoe-shaped' filaments which reveal a {\it laminar} flow pattern
in the gas. By analogy with studies of rising cap bubbles in water, we
assume that the Reynolds number of the flow is 1000 or less and thus
deduce that the viscosity is high, with a factor of ten of the
Spitzer-Braginskii value (Fabian et al 2003b). I conclude that the gas
here is {\it not} turbulent, at least on scales less than 50~kpc.

Studies of flows within the ICM will be opened up by ASTRO-E2 which is
due to launch next February (2005). Its 7~eV spectral resolution will
enable velocities to be determined from Fe K$\alpha$ lines to a few
10s$\kmps$.

Do the likely tangled magnetic fields in the hot gas mean that the
viscosity is much reduced? In accretion discs, the shear in the plasma
increases the viscosity by many orders of magnitude. The situation for
the intracluster medium is unclear but note that Faraday Rotation
studies suggest the field is organised within cells of a few kpc
so approximating a simple field structure on scales comparable to the
ion mean free path. This may mean that the ion viscosity is close to the
Spitzer-Braginskii value. Unlike electron conduction, where
the heat flow can be impeded by mirroring of particles or by an
interface of magnetic structures, sound energy flows
approximately isotropically through such barriers.

\begin{figure}
\includegraphics[width=0.5\textwidth]{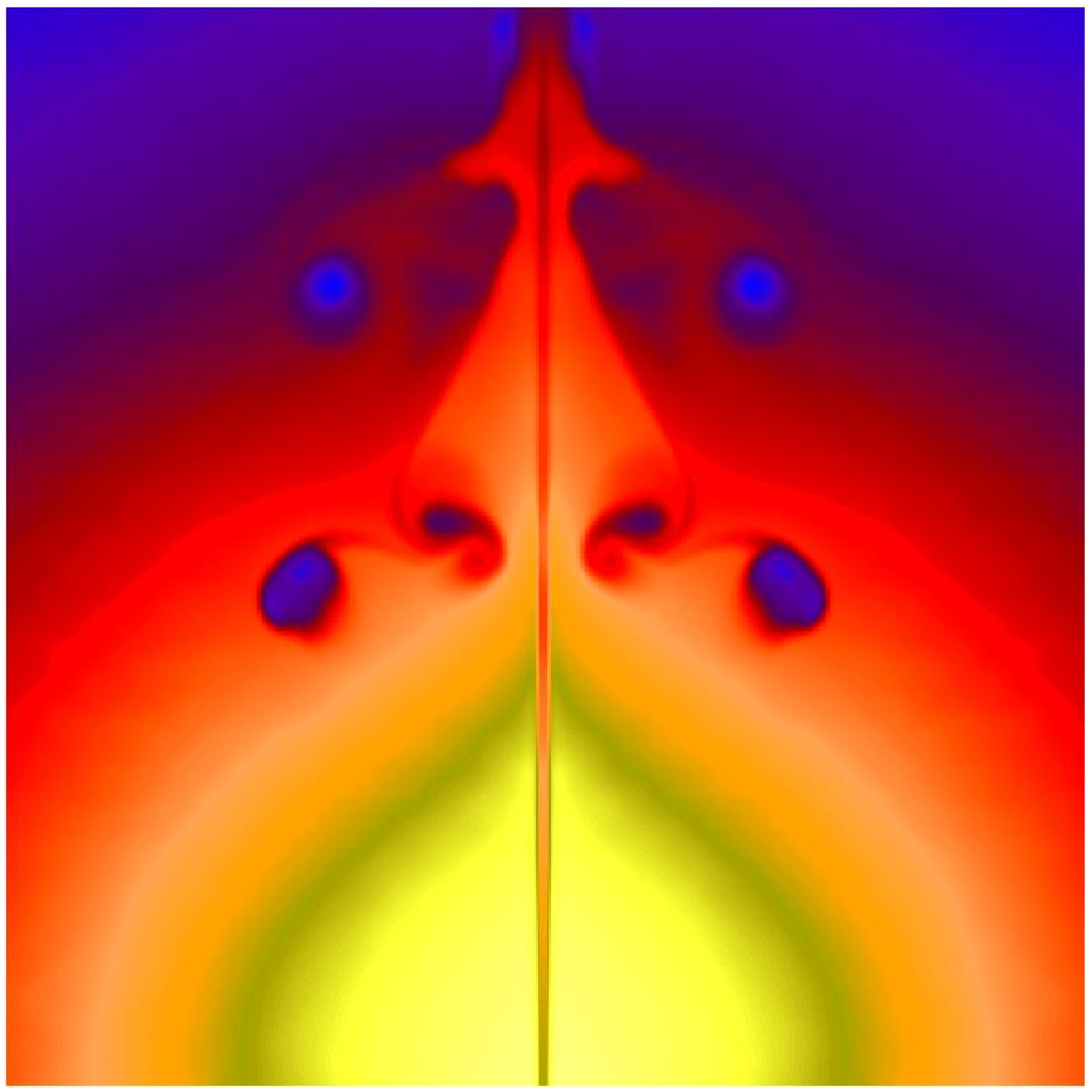}
\includegraphics[width=0.5\textwidth]{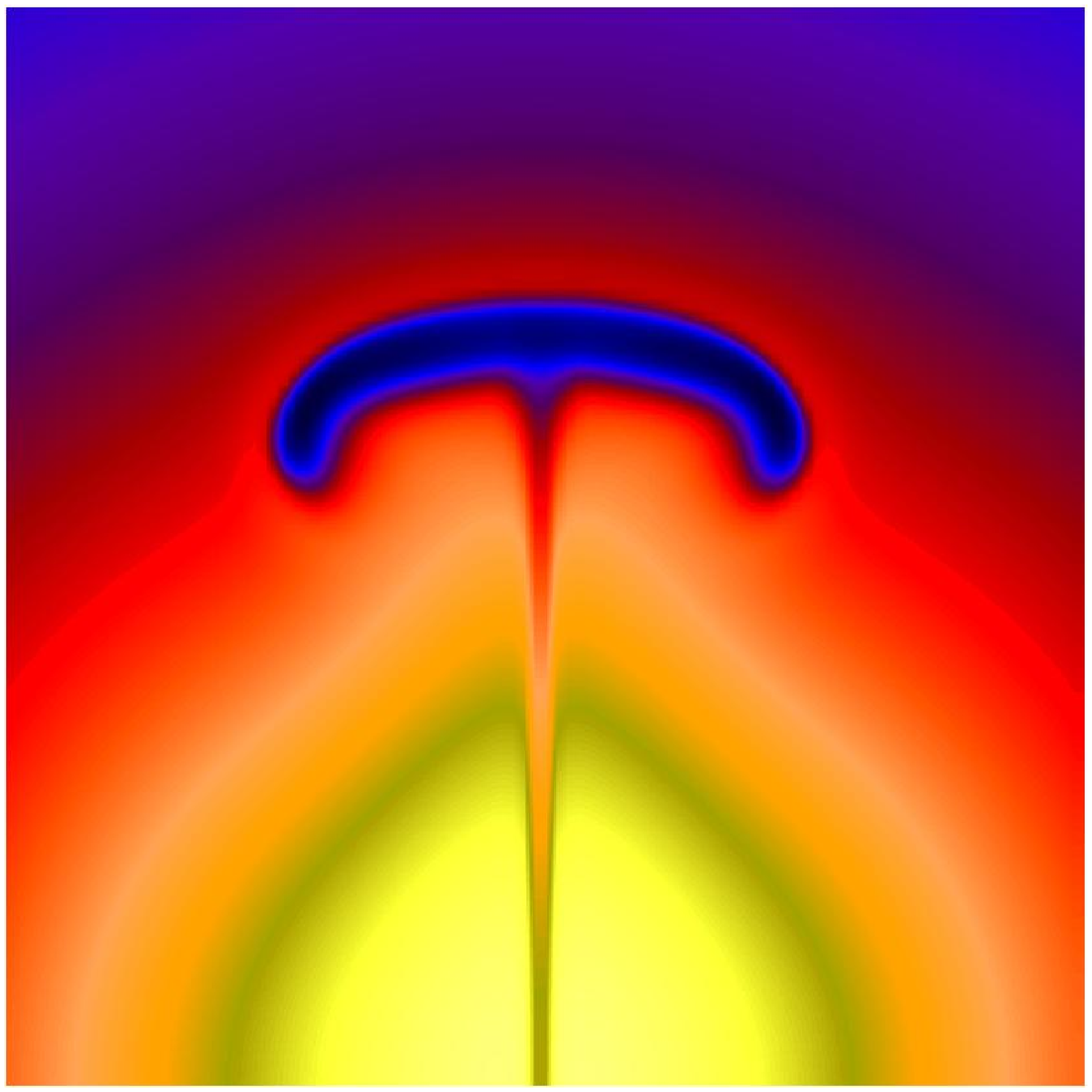}
\caption{Gas density after a bubble has been released (left) in gas
with no viscosity and (right) when the viscosity is one half the
Spitzer value (from Reynolds et al 2004). Note that only in the
viscous case does an intact outer bubble, such as observed in the
Perseus cluster (Fig.~9) occur.}
\end{figure}

Simulations are only just beginning to include viscous effects.
Ruszkowski \& Begelman (2003) and Reynolds et al (2004) have shown
that it can have an important effect on ICM heating due to bubbles. In
particular, the latter study shows that only with significant
viscosity do whole detached bubbles occur (Fig.~10). A viscous medium
can also tap energy from waves generated by subcluster mergers as well
as by a central radio source.

\subsection{General comments}

It is unlikely that either radio source heating or conduction can
suppress radiative cooling within such a large region completely.
Indeed, this is unnecessary, for significant rates of massive star
formation (Crawford et al 1999) and
masses of cold gas (Edge 2001) are found in the central parts. It is
probable that
\begin{equation}
\dot M_{\rm cool}=\dot M_{\rm X}/10,
\end{equation}
where $\dot M_{\rm X}$ is the simple cooling rate derived simply from
the X-ray data on the assumption of no heating. (It is simple to
reduce the rate in equation 3.1 by a factor of 2--3 by including the
gravitational heating and compression of magnetic field in the flow --
the difficulty lies in reducing it by a further factor of 2--3.) The
mass cooling rates previously inferred to be in the range
$100-1000\Msunpyr$ spread over 100~kpc are more likely to be
$10-100\Msunpyr$ concentrated within the inner 30~kpc. The residual
X-ray spectral mass cooling rates are then consistent with the
observed star formation (e.g. RX\,J0820.9+0752, Bayer-Kim et al 2002;
A1068, McNamara, Wise \& Murray 2004; both objects have central radio
sources which are currently weak). Some form of feedback presumably
operates to keep $\dot M_{\rm cool}$ at a low value.

In summary there are plausible heat sources at the centre and beyond
radii of 100~kpc. The main problem is to distribute the energy within
100~kpc without either disrupting the metallicity profiles or
exceeding some observational constraint. Beyond any bubbles and
plumes, and an occasional cold front (all of which occupy only a small
fraction of the volume of the cooler gas) the distribution of surface
brightness, temperature, metallicity and entropy of the gas all vary
very smoothly. {\it Gentle } transport processes in the 
(magnetized) intracluster
gas seem to be crucial. Improvement of our understanding of these
processes is vital for further progress.

\begin{figure}[h]
\begin{center}
\includegraphics[width=0.45\textwidth,angle=-90]{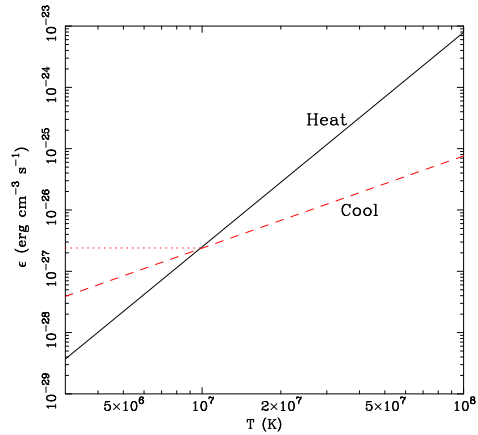}
\caption{Viscous heating rate compared with the cooling rate for an
atmosphere of temperature $T$. Note that heating wins above about
1~keV and cooling below. Sound waves with a wavelength of 10~kpc and
high amplitude have been assumed. }
\end{center}
\end{figure}

Further possibilities remain in which cooling dominates but the
situation is more complicated so that gas cooling below say 2~keV is
less observable. This can result if the metals are not uniformly mixed
in the hot gas (Morris \& Fabian 2003) or if the cooler gas mixes with
cold gas (Fabian et al 2002). The missing soft X-ray luminosity from a
simple cooling flow is similar to that in the optical/UV/IR nebulosity
at the centre. One reason to continue to consider such models is the
detection of strong OVI emission in some clusters (Oegerle et al 2001).

The cooling flow problem, as it has become known, has wider relevance
than just to cluster cores. It is also seen in the central
interstellar medium of elliptical galaxiesm meaning that the problem
occurs over a factor $10^5$ in luminosity! Moreover, 
galaxies grow as gas cools in their dark matter potential wells (White
\& Rees 1978, Kauffmann et al 1999) and the cores of clusters are a
directly observable example of this process. If it does not operate in
cluster cores why does it work in galaxies? It is possible that
whatever is stemming cooling in clusters is less effective in
galaxies, operating in such a way as to dominate only in massive
systems. It therefore determines the upper mass limit of visible
galaxies (Binney 2004). Processes like conduction (Fabian et al
2002b), or viscosity, which are more effective in hotter, massive
objects, have the right property to allow gas to cool in normal
galaxies but not in more massive systems (Fig.~11).

Both turbulent conduction and viscosity are highly temperature
dependent. Kinetic energy fed into the core either by turbulence or
sound waves, generated either from the outer parts of the cluster by
subcluster mergers or from the centre by the central active nucleus,
will dissipate and heat the inner regions, so offsetting the slower
radiative cooling of gas above $10^7\K$. Perhaps the crucial
ingredient is that a massive galaxy has the extensive dense hot
atmosphere at the centre of a group or cluster.

\section{Final comments}

AGN can have, and probably did have, a profound effect on the ICM. The
radio luminosity function (Fig.~1) shows that FRII radio sources
dominated the energy budget at redshifts of 1--3 whereas FRI sources
dominate now. FRII sources acting within the subclusters which have
now assembled into present-day clusters would have seriously heated
and disturbed the ICM there, and could have injected the heat needed
to make $L_{\rm X}\propto T^3$. It is likely that FRI sources in
central cluster galaxies at the present epoch continue to heat the
innermost 100~kpc or so of the ICM gently, preventing central galaxies
from accreting much more gas and determining their final stellar mass.

\section{Acknowledgments}
This is an expanded and updated version of an earlier review. I thank
the Royal Society for support.

\bibliographystyle{mnras}

\label{lastpage}

\end{document}